# Generation of concept-representative symbols
## Position paper

João Miguel Cunha, Pedro Martins, Amílcar Cardoso, Penousal Machado

Departamento de Engenharia Informtica da Universidade de Coimbra
{jmacunha,pjmm,amilcar,machado}@dei.uc.pt

**Abstract.** The visual representation of concepts or ideas through the use of simple shapes has always been explored in the history of Humanity, and it is believed to be the origin of writing. We focus on computational generation of visual symbols to represent concepts. We aim to develop a system that uses background knowledge about the world to find connections among concepts, with the goal of generating symbols for a given concept. We are also interested in exploring the system as an approach to visual dissociation and visual conceptual blending. This has a great potential in the area of Graphic Design as a tool to both stimulate creativity and aid in brainstorming in projects such as logo, pictogram or signage design.

## 1 Introduction

Creativity can be seen as the ability to create novel ideas by making connections between existing ones. It plays an important role in the area of Graphic Design not only in conceiving new concepts but also in visually representing them.

As far as visual representation of concepts is concerned, humans have been doing it since more than two hundred thousand years ago – take for example cave paintings. These representations vary from being completely pictorial – e.g. pictograms – to more abstract – e.g. ideographs.

The link between the visual representation and the conceptual connections behind it can in fact be observed. Examples of this can be seen by looking at Chinese characters, more specifically at the ones categorised as Ideogrammic Compounds (see Figure 1). These characters can be decomposed into others, whose concepts are semantically related, belonging to the same (or at least similar) conceptual space [6].

Some authors were inspired by this relationship between concepts to their visual representations. One of them was Charles Bliss who developed a communication system composed of several hundreds of ideographs that can be combined to make new ones – *Blissymbols* [1]. In his system the variation in terms of abstraction degree can also be observed (see Figure 2).

本　木　林　森

**Fig. 1.** Chinese characters for *root*, *tree*, *woods* and *forest* (left to right). *Root* can be obtained by adding a line to the *tree* character; *woods* character can be obtained (barely) by using two *tree* characters; *woods* can be obtained by using three *tree* characters.

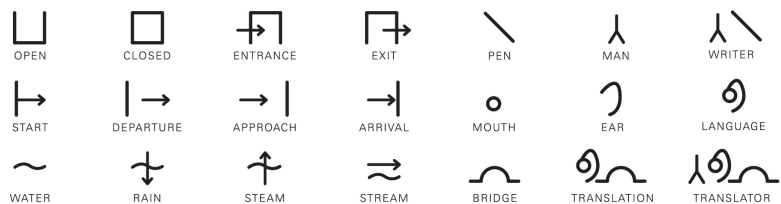

**Fig. 2.** Blissymbols. Several interesting things can be observed by looking at blissymbols: such as a variation in terms of abstraction degree (there are both pictorial and abstract symbols); by combining symbols, new meanings are obtained (examples in Figure 2: *pen + man = writer*, *mouth + ear = language*); by using the same symbols in a different position, a new meaning is obtained (see symbols *water/rain/steam/stream*).

Inspired by examples as the ones presented above, our goal is to conceive an approach for computationally generating concept-representative symbols, i.e. visual representations of concepts. In this paper we present some of the key aspects that have to be considered when generating such symbols and the strategies to explore in order to achieve our objective.

## 2  Generation of concept-representative symbols

The idea of creating a symbol for a given concept based on its connections to other concepts is, just by itself, interesting. However, if we consider the exploration of this idea using computational means to automate the generation process of the symbols, its potential greatly increases.

We can think of a tool capable of generating symbols whose visual properties would be the outcome of an analysis of the conceptual space of the introduced concept. We believe that such a tool could assist the designer during the ideation process by stimulating its creativity, aiding in brainstorming activities and thus giving rise to new ideas and concepts.

Concerning the visual qualities of the generated symbols, it is crucial to consider several aspects. The first one is the degree of abstraction. This aspect can be considered to be influenced by the choice of the connections used in the symbol generation. Take for example the concept *car*: if we consider the connections between *car* and the concepts *door*, *window* and *wheel*, the resultant symbol will probably be highly pictorial; if we choose to ignore those connections,

158

the resulting symbol might be more abstract. Ideally, the tool should allow the abstraction degree to be set according to the user's needs.

However, this is not the only aspect that is greatly dependent on the connections used. As observed in the blissymbols, a given combination of symbols might lead to different interpretations. If some perceptual aspects are not considered, this might result in a conflict between the concept and the perception of the symbol. In the next subsection some of these aspects will be presented.

### 2.1 Considering visual characteristics

When dealing with symbols, it is important to bear in mind some aspects of how a representation's meaning can be changed by changing some of its visual characteristics. The following aspects are essential: position, colour and shape.

The first semiotic aspect – position – can be seen in Figure 2. By putting an arrow next to the *water* symbol a new concept is represented. The concept also varies according to positioning of the arrow. Such details must be considered and a mechanism for analysing them needs to be developed (similar issues have been considered in [2]).

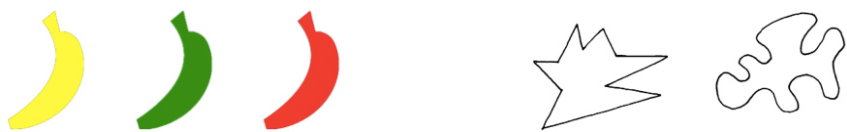

**Fig. 3.** <u>Left side</u> is shown how the meaning of a banana can change with its colour (mature, green and red banana). <u>Right side</u> is Kiki/Bouba example. Accordingly to Ramachandran 98% of all respondents atribute the name Kiki to the shape on the left and Bouba to the right one, despite having no meaning at all [9]. *Best viewed in colour.*

Another aspect to be regarded is colour. Through a brief analysis of the banana example in Figure 3 it is easy to understand the importance of this aspect. By simply attributing a different colour to the same symbol, its perceptual meaning also changes. In addition, the use of colour has already been proven as a mean of facilitating the interpretation of visual representations of concepts (e.g. [7]). However, its incorrect use has the opposite effect (e.g. Stroop effect), causing interference in its interpretation. On the other hand, a mechanism to avoid an over-use of colour will probably be needed as colour might not be necessary in some symbols.

The third important aspect is shape. When generating visual symbols from textual data (e.g. semantic networks), one cannot avoid dealing with shape. The choice of shape for a given concept is not easy by itself but one has also to consider its visual qualities. Take for example the shapes presented in Kiki/bouba example in Figure 3. Despite not having any meaning at all, there is a clear tendency or bias when attributing names to them. This can be explained as follows, humans tend to perform mappings among domains, namely between image and sound, as such sharp shapes tend to be associated with sharp sounds and organic shapes with smooth ones [9].

As we have already mentioned, these semiotic aspects have to be considered when generating symbols. This is only possible to achieve by thoroughly analysing the conceptual network and also considering previously generated symbols as both examples and base for the generation of the new symbol.

### 2.2  Getting information

An extensive analysis of the conceptual space is important but there is another issue that has to be resolved: if the system does not have access to a large source of knowledge – with information about visual properties – it will be difficult, if not impossible, to achieve good results. One possibility is to use an already built semantic network of common sense knowledge (e.g.[8] or [3]).

However, as our main objective is to generate visual representations, knowledge about visual characteristics is required. For that reason, a methodology has to be conceived for acquiring such data. A possible solution is to use a similar approach to the one used by Open Mind Common Sense project – a knowledge acquisition system designed to acquire common sense knowledge from the general public over the web [10]. Our goal is to focus on gathering information about objects' visual characteristics such as colour, shape and texture. These will likely allow us to attain adequate results in terms of symbol generation.

Crowdsourcing will probably be used in our knowledge-gathering process as it easily allows to reach a high number of contributors at a reduced cost. In addition, the validity of online crowdsourced experiments on visual properties and graphical perception has already been demonstrated (e.g. [5]).

This *distributed human project* approach allows us not only to gather data at a scale that would not be possible otherwise but also enables us to study the role of context in perception – one of our goals is to test whether the symbols generated differ accordingly to the location where the data was gathered from.

We also intend to explore other alternatives for populating our semantic network, such as automatic gathering of information. Using Google Image Search is one example of this and can be used to find images related to the content being analysed and consequently extracting useful information from them (e.g. [7]).

### 2.3  Generating symbols

In our opinion there are, at least, two different ways of generating a symbol for a given concept: (1) starting with no prior knowledge and analysing the conceptual space in order to extract possible visual features to be used in the symbol; (2) using prior knowledge – previously generated symbols of concepts that are, in some way, related to the introduced concept. This would lead to a higher coherence among generated symbols. In both cases, not only direct conceptual connections are used but also more uncommon ones – through a process of analogy. As such, we argue that mechanisms such as case-based reasoning or conceptual blending [4] are suitable strategies to generate symbols of concepts. As for the former, we can consider a case-base comprised of symbols such as the ones depicted in Figure 2 and develop a system to produce novel ones using analogues to the previous ones. Regarding conceptual blending, the idea is to explore the structure mapping approach to analogy and concept integration based on conceptual spaces and semiotic systems.

## 3   Conclusion

In this paper we presented our approach to computational generation of concept-representative symbols. We aim to develop a design aiding tool that combines the exploration of conceptual spaces in combination with processes of analogy-making and semiotic analysis to generate possible visual (abstract/semi-abstract) representations for the concepts introduced by the user. We believe that it will help the designer during the ideation process by stimulating its creativity, aiding in brainstorming activities and thus giving rise to new ideas and concepts.

**Acknowledgements**

The authors acknowledge the financial support from the Future and Emerging Technologies (FET) programme within the Seventh Framework Programme for Research of the European Commission, under the ConCreTe FET-Open project (grant number 611733) and the PROSECCO FET-Proactive project (grant number 600653).